\documentclass[fleqn,usenatbib]{mnras}

\usepackage{newtxtext,newtxmath}

\usepackage[T1]{fontenc}

\DeclareRobustCommand{\VAN}[3]{#2}
\let\VANthebibliography\thebibliography
\def\thebibliography{\DeclareRobustCommand{\VAN}[3]{##3}\VANthebibliography}

\usepackage{graphicx}	
\usepackage{amsmath}	
\usepackage{multirow} 	

\hypersetup{pdfauthor={M. Sarkis},
pdftitle={Gamma Rays from UltraCompact Minihaloes: Effects on the Earth's Atmosphere and Links to Mass Extinction Events},
pdfkeywords={gamma-rays: general -- atmospheric effects -- dark matter},
bookmarksnumbered=true}

\title[Atmospheric effects from UCMHs]{Gamma Rays from UltraCompact Minihaloes: Effects on the Earth's Atmosphere and Links to Mass Extinction Events}

\author[M. Sarkis, G. Beck \& B. C. Thomas]{
M. Sarkis,$^{1}$\thanks{E-mail: michael.sarkis@students.wits.ac.za}
G. Beck$^{1}$
and B. C. Thomas$^{2}$
\\
$^{1}$Department of Physics, University of the Witwatersrand, Private Bag 3, WITS-2050, Johannesburg, South Africa\\
$^{2}$Department of Physics and Astronomy, Washburn University, Topeka, KS 66621, USA\\
}

\date{Accepted XXX. Received YYY; in original form ZZZ}

\pubyear{2020}

\begin{document}
\label{firstpage}
\pagerange{\pageref{firstpage}--\pageref{lastpage}}
\maketitle

\begin{abstract}
Recent studies of the effects on the Earth's atmosphere by astrophysical sources, such as nearby
gamma-ray bursts or supernovae, have shown that these events could lead to severe changes
in atmospheric composition. Depletion of ozone, the most notable of these changes, is extremely dangerous to living organisms as any decrease in ozone levels leads to an increase in the irradiance of harmful solar radiation at the Earth's surface. In this work we consider dark matter as an astrophysical source of gamma rays, by the annihilation and decay of WIMPs found within dark compact halo objects known as UltraCompact Minihaloes (UCMHs). We calculate the fluence of gamma rays produced in this way and simulate the resulting changes to terrestrial ozone levels using the Goddard Space Flight Center 2D Atmospheric Model. We also calculate the rate at which such events would occur, using estimates for the mass distribution of these haloes within the Milky Way. We find that the ozone depletion from UCMHs can be significant, and even of similar magnitude to the levels which have been linked to the cause of the Late-Ordovician mass extinction event. However, the probability of such encounters over the Earth's entire history is relatively low. This suggests that, while dark compact objects like UCMHs could have had an impact on the Earth's biosphere, other astrophysical phenomena like gamma-ray bursts or supernovae seem a more likely source of these effects. 
\end{abstract}




\section{Introduction} \label{sec:intro}

Astrophysical phenomena have often been investigated as the underlying causes for past mass extinction events. Research into the terrestrial effects of gamma-ray bursts (GRBs) and supernovae (SN) ejecta using detailed atmospheric models and simulations~\citep{ellis1995,crutzen1996,gehrels2003,thomas2005,ejzak2007,thomas2015,thomas2018SN,melott2017} have yielded robust evidence of the substantial biological impact caused by these phenomena. The most notable effect of an ionizing flux on the atmosphere, first investigated by~\citet{ruderman1974}, is the depletion of ozone molecules found predominantly in the Earth's stratosphere. This relatively small layer of ozone effectively absorbs solar ultraviolet radiation, and a depleted ozone layer would lead to living organisms on the surface being exposed to intense and harmful levels of solar UV. If important primary producers like phytoplankton are significantly affected, the global food chain could face a crisis. The biological consequences of a depleted ozone layer were shown to be consistent with the fossil record of the late Ordovician period $\sim443$ million years ago, during which there were major mass extinction events, and the source of the ozone depletion in this case was attributed to a GRB impacting the atmosphere at latitudes south of $-75^{\circ}$~\citep{melott2009}. Further studies have shown that nearby SN are also capable of significant ozone depletion, and could have contributed to biodiversification of species around the end of the Pliocene epoch $\sim2.5$ million years ago~\citep{thomas2016SN,thomas2018SN}. While these studies do not confirm GRBs or SN as causes of mass extinctions, they do suggest that these types of astrophysical phenomena are capable of having a significant impact on our biosphere. 

Dark Matter (DM) has also been considered as the culprit for past mass extinction events on the Earth. The most popular phenomenon studied in the literature is of gravitational perturbations of objects in the Oort cloud from a disk of DM coincident with the Milky Way's midplane, referred to as a dark disk~\citep{randall2014,kramer2016}. The existence of such a disk has yet to be confirmed, but the results of~\citet{kramer2016} show that solar system oscillations through a dark disk are compatible with the crater impact rate observed on Earth. As the traversal of the Earth through this disk would be periodic, this effect could also partly explain the believed periodicity present in mass extinction events~\citep{raup1984,rohde2005} -- a factor that transient phenomena like GRBs cannot. Another interesting link between DM and mass extinction events is based on the capture of weakly interacting massive particles (WIMPs) by the Earth's gravitational field. If enough WIMPs were captured for an appreciable amount of annihilation to occur, the collision of annihilation products with nuclei in the Earth's core could lead to a substantial amount of internal heating~\citep{abbas1996,rampino2015}. The raised temperature of the core and mantle would then eventually result in large scale flood-basalt volcanism on the surface, a phenomenon which has already been associated with the Permian-Triassic mass extinction event~\citep{campbell1992}. The final noteworthy link between DM and terrestrial biodiversity involves the direct and mutagenic collisions between WIMPs and DNA molecules. It has been argued by~\citet{collar1996} that the traversal of the Earth through a region of high DM density would lead to a higher rate of these collisions, which could result in widespread carcinogenesis among all living organisms. However, it was recently shown that with more accurate organic tissue modelling and better WIMP-nuclei cross section data the results of~\citet{collar1996} are likely overestimated~\citep{freese2012,sarkis2019}. Although this mechanism has the potential to affect living organisms in a very direct way, the consequences to global populations currently seem negligible.

We now consider a new astrophysical mechanism with potential biological impact: the interaction of WIMP annihilation and decay products with the Earth's atmosphere. We focus on the particular case of gamma-ray photon final states, calculated here using the numerical results of the PPPC4DMID package~\citep{cirelli2011}, and simulate the atmospheric effects using the Goddard Space Flight Centre (GSFC) 2D atmospheric model. The primary result of this study is the likelihood and extent of ozone depletion in the atmosphere, which allows for comparison to other results that consider GRBs as ionization sources. Importantly, we also consider `clumpy' regions of DM substructure in the Milky Way, the presence of which is heavily suggested by numerical DM simulations~\citep{kuhlen2012}. We specifically focus on the case of UltraCompact Minihaloes (UCMHs), recently proposed dark compact halo objects with characteristically steep density profiles~\citep{ricotti2009}. As the flux of gamma rays from a DM source is strongly dependent on the density of WIMPs in that region, these dense objects can significantly augment the mechanism of ozone depletion investigated here. We place a rough estimate on the mass distribution of these UCMHs from constraints on primordial black hole (PBH) distributions, which is motivated by the similar formation scenarios of these objects, and use this to determine the probability of a deadly encounter with an UCMH over the Earth's history. The structure of this paper is as follows: in Section~\ref{sec:dm} an outline of the DM models used in this work is provided. Section~\ref{sec:atmospheric_modelling} contains a description of the effects of ionizing radiation on the atmosphere and the potential biological effects associated with these changes. The results are presented and discussed in Section~\ref{sec:results} and a summary of the results is given in the conclusion in Section~\ref{sec:conclusion}.

\section{Gamma rays from UCMHs}\label{sec:dm}

\subsection{DM halo models}\label{sec:halos}
The internal structure of DM haloes have been probed with N-body simulations in the past decades~\citep{navarro2010}, with the prominent Navarro-Frenk-White (NFW)~\citep{nfw1997}, Einasto~\citep{einasto} and Burkert~\citep{burkert1995} halo density profiles showing good fits to simulated DM haloes. There is also evidence from high-resolution simulations that more compact clumps of DM, so-called subhaloes, can exist within the larger parent haloes as a result of the hierarchical formation of DM structures~\citep{belikov2012}. The increased density of DM within these subhaloes leads to more WIMP annihilations and a larger flux of observable annihilation products. One halo density profile of interest in this regard is of the recently hypothesised UCMH, which was originally proposed as a dark compact halo object that could contribute to the microlensing signals in searches for MACHOs~\citep{ricotti2009}. The formation of these haloes would be seeded by the gravitational collapse of over-densities present in the radiation-dominated epoch, after which they would grow in mass by the accretion of surrounding matter. The amplitude of the density perturbations ($\delta$) needed to form UCMHs would be smaller than the $\delta \sim 0.3$ threshold necessary to form primordial black holes (PBHs), and can be as small as $\delta \sim 10^{-3}$~\citep{berezinsky2003}. One implication of this formation scenario would be that UCMHs are likely more abundant than PBHs, which are already tightly constrained DM candidates~\citep{josan2009,lacki2010}. From principles of spherical collapse and radial infall, the redshift-dependent radial density profile of an UCMH can be written as
\begin{equation}\label{eqn:ucmh_density}
\rho_{\text{UCMH}} (r,z) = \dfrac{3f_{\chi}}{16\pi}\dfrac{M_{\text{UCMH}}(z)}{{R_{\text{UCMH}}(z)}^{\frac{3}{4}}}\dfrac{1}{r^{\frac{9}{4}}} \,,
\end{equation}
where $M_{\text{UCMH}}$ and $R_{\text{UCMH}}$ are the virial mass and radius of the UCMH. The value of $f_{\chi} = \Omega_{\chi}/\Omega_{m}$ represents the fraction of total matter in the universe which is dark, as the UCMH seed would grow from the accretion of both DM and baryonic matter after the time of recombination. The radius $R_{\text{UCMH}}$ of the halo is  
\begin{equation}\label{eqn:radius}
R_{\text{UCMH}}(z) = 0.019\left(\dfrac{1000}{z+1}\right)\left(\dfrac{M_{\text{UCMH}}(z)}{M_{\odot}}\right)^{\frac{1}{3}}\, \text{pc}.
\end{equation}
We consider the growth of UCMHs to continue until the epoch of star formation at $z \sim 10$, after which the accretion of background matter is expected to be inefficient~\citep{bringmann2013}. Although the profile given in Equation~\ref{eqn:ucmh_density} has a steep radial dependence, we also expect the density in the inner regions of the halo to be reduced by the annihilation of WIMPs. This process provides an upper limit on the density inside the halo, which has been estimated by~\citet{ullio2008} as
\begin{equation}\label{eqn:rhomax}
\rho_{\text{max}} \simeq \frac{m_{\chi}}{\langle \sigma v \rangle (t_0 - t_c)}\, .
\end{equation}
Here the value $m_{\chi}$ is the mass of an individual WIMP, $\langle \sigma v \rangle$ represents the velocity-averaged cross section for WIMP annihilations and $(t_0 - t_c)$ is the age of the halo since collapse. We thus take the density of the halo to be the minimum value between Equation~\ref{eqn:ucmh_density} and Equation~\ref{eqn:rhomax} for each radius inside the halo. 

Doubts surrounding the exact form of the density profile of these haloes, especially within the inner regions, have emerged recently through the work of~\citet{gosenca2017}. In this study it was shown that while the UCMH profile can be reproduced in an independent simulation, the formation of such a halo is unlikely given a more realistic formation scenario. This scenario includes initial overdensity seeds that do not form in isolation from other perturbations of similar scale as well as the introduction of 3-dimensional effects (such as angular momentum) within the halo. These results have been supported in another study by~\citet{delos2018_1}, wherein it is shown that haloes with a shallower profile than the steep $\rho \propto r^{-9/4}$ dependence are favoured throughout the halo's formation. We therefore introduce a new halo profile for comparative purposes, given by 
\begin{equation}\label{eqn:moore_density}
\rho_{\text{MOORE}} (r) = \dfrac{\rho_{s}}{(r/r_s)^{3/2}(1+r/r_s)^{3/2}}  \,.
\end{equation}
Here $\rho_s$ is a characteristic density necessary for normalisation, and $r_s$ a scale radius. This is the profile first introduced by~\citet{moore1999} (called \textit{Moore} or \textit{Moore-like} from now) which fits the simulated minihaloes in~\citet{delos2018_2} well. The density of this halo profile is lower than in UCMHs within the inner regions of the halo, which has a significant effect on output annihilation fluxes because of the steep nature of the UCMH profile. We do however limit the density of both halo profiles with the value calculated in Equation~\ref{eqn:rhomax}. 

\subsection{WIMP annihilation and decay}\label{sec:annihilation}
The annihilation and decay of WIMPs inside the DM clumps is expected to produce a spectrum of SM particles that travel toward and interact with the Earth's atmosphere.  In this work we focus on the prompt emission of gamma rays, with the instantaneous flux of these particles given by the following equations:
\begin{equation}\label{eqn:flux}
  \frac{\mathrm{d} \Phi}{\mathrm{d} E} = 
    \begin{cases}  
    \dfrac{\langle \sigma v \rangle}{8\pi m_{\chi}^2}\, J\, \sum_f \dfrac{\mathrm{d}N_{\gamma}^f}{\mathrm{d}E}B_f  \,, & \text{(annihilation)} \\[1.5em]
    \dfrac{\Gamma_{\text{dec}}}{4\pi m_{\chi}}\, D\, \sum_f \dfrac{\mathrm{d}N_{\gamma}^f}{\mathrm{d}E}B_f   \,, & \text{(decay)}
  \end{cases}
\end{equation}
where $\Gamma_{\text{dec}}$ is the decay rate of an individual WIMP, and we have assumed that WIMPs are self-conjugated for annihilation (were this to not be true, the observed flux would be further reduced by a factor of 2). The sum runs over index $f$, which represents the unique intermediate interactions in the annihilation/decay process, known as channels. The quantity $B_f$ is the branching ratio for each channel, and we consider particular channels individually, performing calculations with $B_f = 1$ for each channel of interest. In this work we focus on representative spectra with large relative output fluxes, finding results primarily for the $q\bar{q}$ channel which consists of light quark-antiquark pairs. The spectra $\mathrm{d}N_{\gamma}^f/\mathrm{d}E$ are typically computed numerically through the use of software packages, and here we use the spectra from the PPPC4DMID package provided in~\citet{cirelli2011}. These results only include prompt emission of gamma-rays without any secondary emissions from bremsstrahlung or inverse Compton processes, which is applicable to the scenario studied here since the region of interest is the vicinity of the solar system. The results have also been computed with electroweak and polarisation corrections~\citep{electroweak}, which has been shown to be important for TeV-scale WIMP masses. Recent cosmic ray detections in this $\sim$~TeV mass range allow for an annihilating DM interpretation~\citep{beck2019}, a result which has renewed interest in these high-mass WIMPs. In this work we examine a large portion of the possible WIMP mass window, covering a range of $m_{\chi} = 10$~GeV to $m_{\chi} = 1$~TeV. Finally, the quantities $J$ and $D$ in the expression for the flux, known as the `astrophysical $J$/$D$ factors', encapsulate the distribution of DM in the region of interest and are computed as follows: 
\begin{equation}\label{eqn:jfactor}
  J = \displaystyle{\int_0^{\mathrm{r_{95}}}}  \dfrac{\rho^2(r)\,r^2}{(r^2 + d^2)} \, \mathrm{d}r \, ,
\end{equation}
and
\begin{equation}\label{eqn:dfactor}
  D = \displaystyle{\int_0^{\mathrm{r_{95}}}}  \dfrac{\rho(r)\,r^2}{(r^2 + d^2)} \, \mathrm{d}r \, .  
\end{equation}
In these equations $r$ is the radius of the spherically symmetric halo, $d$ is the distance from the Earth to the centre of the halo and r$_{95}$ is a radius that contains $\sim 95$ per cent of the total gamma-ray flux from the halo. The $J$ and $D$ factors as presented above account for the extension of the halo's radius, as most of the haloes considered here appear as extended sources due to their proximity. The angular size of each halo is then determined using the formula
\begin{equation}\label{eqn:halo_size}
\delta = 2\arctan\left(\dfrac{\mathrm{r_{95}}}{d}\right) \,.
\end{equation}

It should finally be noted that the steep radial dependencies of each density profile leads to a majority contribution to the output flux from the central regions of the halo. This has direct consequences for the ozone depletion results at various WIMP masses, as the maximum density of the halo (given by Equation~\ref{eqn:rhomax}) is directly proportional to the WIMP mass. \\
When considering the interaction of gamma rays on the Earth's atmosphere, a study by~\citet{ejzak2007} has shown that the total gamma-ray fluence, rather than the instantaneous flux (as calculated in Equation~\ref{eqn:flux}), is a more useful tracer for ozone depletion. In this study long-term effects on the atmosphere from incident ionizing radiation do not show any significant changes when the time-scale of energy deposition into the atmosphere is in the range of $0.1-10^8$~s. The time intervals considered here fall well within this range, which motivates the following calculation of the total fluence ($F$) that will be used as an input for the atmospheric model used in this work. We thus calculate $F$ by simply integrating over the time interval of interest ($\Delta t$) as well as the energy range of the spectrum $\mathrm{d}N_{\gamma}/\mathrm{d}E$, i.e. 
\begin{equation}\label{eqn:fluence}
F = \int_0^{\Delta t} \int_{E_{\text{min}}}^{E_{\text{max}}}\, \left(\dfrac{\mathrm{d}\Phi}{\mathrm{d}E}(t) E\right) \mathrm{d}E \mathrm{d}t \,.
\end{equation}
The implicit time-dependence of the instantaneous flux follows from the variation in relative distance to each UCMH/Moore-like halo as it moves through its orbit around the galactic centre. We assume for simplicity that the apparent trajectory of each halo is a linear path over the considered time interval, and determine the distance $d$ to the halo at some time $t$ as
\begin{equation}\label{eqn:distance}
d(t) = \sqrt{b^2 + (v_{\mathrm{H}}t)^2} \,.
\end{equation}
Here $v_{\mathrm{H}}$ is the velocity dispersion of the halo and $b$ is a chosen impact parameter, which is the minimum perpendicular distance between the centre of the halo and the Earth at any point. 

\subsection{Mass distribution and encounter rate}\label{sec:encounters}
The mass distribution function, denoted here by $\Psi(M)$, is an important factor when determining the number of encounters the Earth would have with DM clumps in its orbit around the galaxy. Since the formation of UCMHs is believed to be similar to that of PBHs, we adopt an approach of estimating the mass distribution of UCMHs that mirrors the current estimates for the PBH distribution. The current constraints on the PBH distribution come from several observations, most notably from microlensing surveys and measurements of dynamical effects of objects in the Milky Way~\citep{green2014}, and in this work we assume $\Psi(M)$ has the following log-normal form:
\begin{equation}\label{eqn:mass_distribution}
\Psi (M, \sigma, \mu) = \dfrac{1}{\sqrt{2\pi}\sigma} \exp{\left(-\dfrac{\log^2(M/\mu)}{2\sigma^2}\right)} \,,
\end{equation} 
where $(\sigma, \mu)$ are the usual parameters that determine the shape of the distribution. Additionally, for any values of $(\sigma, \mu)$ the following normalisation condition is enforced:
\begin{equation}\label{eqn:distribution_integral}
\int_0^{\infty} \dfrac{\Psi (M)}{M} \mathrm{d}M = 1 \,.
\end{equation}
This distribution was chosen in line with the analytic derivation of the mass-distribution for PBHs and MACHOs originally presented in~\citet{dolgov1993}, which stemmed from the evolution of quantum fluctuations present during inflation. This has since been supported by further studies~\citep{dolgov2009,blinnikov2016}, which suggests that the distribution in Equation~\ref{eqn:mass_distribution} is a suitable estimate for the dense DM subhaloes considered in this work. The calculation of the rate of encounters with DM clumps experienced by the Earth is then given by 
\begin{equation}\label{eqn:encounter_rate}
\dfrac{\mathrm{d}\Gamma}{\mathrm{d}t} = \pi^2 f_{\text{H}} v_{\text{E}} \rho_0 \int_0^{\infty}\, d_{\text{enc}}^2 \dfrac{\Psi (M)}{M^2} \mathrm{d}M \,. 
\end{equation}
This calculation follows the general procedure laid out in~\citet{green2016} (and references therein) for finding the rate of PBH lensing events. These equations have been modified for the purposes of this work, with $f_{\text{H}}$ representing the fraction of total DM in the galactic halo that is contained in UCMHs or Moore-like clumps, $v_{\text{E}}$ giving the relative orbital velocity of the Earth around the galactic centre to the orbital velocity of an UCMH or Moore-like halo and $\rho_0$ being the DM density at the Earth's position in the absence of any subhaloes. The quantity $d_{\text{enc}}$ represents the maximum distance from the centre of a clump to the Earth that would produce a threshold fluence of gamma rays, usually taken as one that would cause globally significant effects (see Section~\ref{sec:atmospheric_modelling}). From this derivation, any clump that passes within a distance $d_{\text{enc}}$ to the Earth will be considered as an `encounter'. The total number of encounters that take place over a specified time interval is then
\begin{equation}\label{eqn:encounter_number}
\Gamma = \int_0^{\tau} \frac{\mathrm{d}\Gamma}{\mathrm{d}t} \mathrm{d}t \,,
\end{equation}
with $\tau$ as the time interval. In all calculations this is taken to be the age of the Earth, estimated at $\sim 4.5 \times 10^9$ years. Therefore the results shown in this work represent the number of possible encounters with the Earth since its formation. \\

\section{Atmospheric Effects}\label{sec:atmospheric_modelling}

We model all changes to atmospheric chemistry from an incident flux of gamma rays with the GSFC 2D atmospheric model. This is a time-dependent, latitude-altitude model that simulates the chemical reactions of atmospheric constituent species in the presence of solar radiation and cosmic ray backgrounds. The model runs from the ground to 116 km in altitude, with approximately 2 km altitude bins, and from pole-to-pole in 18 bands of $10^{\circ}$ latitude each. The model includes 65 chemical species, 37 transported species and ``families'' (e.g., NO$_y$), winds, small-scale mixing, solar cycle variations, and heterogeneous processes (including surface chemistry on polar stratospheric clouds and sulphate aerosols). We use the model in a pre-industrial state, with anthropogenic compounds (such as CFCs) set to zero. The primary effect of interest here -- the depletion of ozone -- is facilitated by the formation of NO and OH molecules~\citep{ruderman1974,muller1993}. These molecules are able to participate in a series of chemical reactions that catalytically remove ozone from the atmosphere, and are formed in the presence of ionizing radiation in the stratosphere. The modelling of atmospheric effects in this work closely follows the studies performed by~\citet{thomas2005,thomas2015} that investigate the terrestrial effects of ionizing events using the GSFC model. The detailed results presented therein, that rely on the determination of ozone depletion and inferred biological consequences, seem to be applicable to any ionizing event with similar radiation signatures (particularly of fluence and spectral hardness)~\citep{ejzak2007,thomas2015}. Thus, the source of ionization in the atmosphere is the primary difference between this work and the studies mentioned above; here we use the gamma-ray fluxes generated from DM haloes passing nearby the solar system, as opposed to GRBs (or supernovae), to calculate the ionization rate profiles needed as input for the atmospheric model. 

Ionization rate profiles are calculated separately from the atmospheric model, following the method used in~\citet{gehrels2003,thomas2005,ejzak2007}. The total photon flux in each of 181 energy bins in the range $10^{-3}\, \mathrm{MeV} \leq E \leq 10^{6} \,\mathrm{MeV} $, calculated as described in Section~\ref{sec:results_fluxes}, is propagated vertically through a standard atmosphere (adjusted for the appropriate latitude and time when input to the atmospheric model), attenuated with altitude by an exponential decay law with energy-dependent absorption coefficients taken from a lookup table. The lookup table values for $10^{-3}\, \mathrm{MeV} \leq E \leq 10^{5} \,\mathrm{MeV} $ were obtained from the National Institute of Standards and Technology (NIST) XCOM database, available on-line~\citep{berger2005}, based on a mixture of 79 per cent  N$_2$ and 21 per cent  O$_2$.  That database does not extend above $10^{-3}\, \mathrm{MeV}$, so we have generated values $10^{5}\, \mathrm{MeV} \leq E \leq 10^{6}\, \mathrm{MeV} $ using a log-log extrapolation based on the database values between $10^{4}\, \mathrm{MeV}$ and $10^{5}\, \mathrm{MeV}$. The energy deposited in each atmospheric layer is computed and then converted to an ionization rate using 35 eV per ion pair~\citep{porter1976}. The vertical ionization rate profiles are then mapped onto the altitude and latitude grid used by the GSFC model. We now outline the prominent attributes of ozone depletion as calculated by the model, and refer the reader to~\citet{thomas2005} for a description of all the chemical species and reactions present in the model.

\subsection{Ozone depletion}\label{sec:ozone_depletion}

The abundance of atmospheric ozone is believed to be dynamically balanced by the simultaneous production and destruction of ozone molecules, which can occur through multiple reaction chains~\citep{madronich1993}. Production of ozone in the stratosphere occurs via the photodissociation of dioxygen molecules by short-wavelength ($<$ 240 nm) solar UV light, or
\begin{equation}\label{eqn:ozone1}
\mathrm{O_2 + h\nu (\lambda < 240nm) \rightarrow 2 O(^3 P)}\,,
\end{equation}
which leads to the formation of O$_3$ molecules via
\begin{equation}\label{eqn:ozone2}
\mathrm{O(^3 P) + O_2 \rightarrow O_3}\,.
\end{equation}
Reaction~\ref{eqn:ozone2} can also take place in the presence of a 3rd molecule, usually in the form of an abundant atmospheric species such as N$_2$ or O$_2$. Reactions~\ref{eqn:ozone1} and~\ref{eqn:ozone2} constantly compete with a number of reaction chains that destroy ozone molecules, referred to as catalytic cycles. These involve groups of other molecules, most notably parts of the odd oxygen O$_x$ (O, O$_3$), odd nitrogen NO$_x$ (NO, NO$_2$), odd hydrogen OH$_x$ (OH, HO$_2$) and chlorine ClO$_x$ (Cl, ClO) families. A common feature in all of the catalytic cycles is that the constituent molecules survive the reaction, even up to hundreds of times before interacting with other species~\citep{gehrels2003}. An instructive example is of the NO$_x$ cycle, where in the presence of NO and NO$_2$ the following reactions occur
\begin{equation}\label{ozone3}
\mathrm{NO + O_3 \rightarrow O_2 + NO_2} \,,
\end{equation}
and 
\begin{equation}\label{ozone4}
\mathrm{O(^3 P) + NO_2 \rightarrow O_2 + NO}\,.
\end{equation}
As the NO and NO$_2$ molecules survive this reaction chain, they are able to remove ozone from the atmosphere catalytically. While the cycles involving other molecular groups proceed similarly, the odd nitrogen and odd hydrogen cycles are of particular importance here as their constituents are formed in the presence of ionizing radiation. Specifically, in the GSFC model each electron-ion pair formed from an ionizing event is taken to produce 1.25 NO$_x$ molecules and between 0 and 2.00 OH$_x$ molecules depending on the altitude at which the ionization takes place~\citep{porter1976,solomon1981}. The formation of these molecules and the cycles they participate in are the primary mechanisms through which ozone is depleted in this study. 

In addition to the rates of the reactions mentioned above, there are several geographic and seasonal factors that can affect the abundance of ozone at some local area and time. In general, these factors could affect the final ozone depletion levels for various astrophysical phenomena differently, based on the way the radiation enters the atmosphere. Firstly, the presence or absence of sunlight directly affects the rate of ozone formation, as the production of ozone (through Reactions~\ref{eqn:ozone1} and~\ref{eqn:ozone2}) depends critically on the availability of solar radiation for photodissociation of O$_2$. Further, NO$_y$ constituents (which include NO$_x$ and other nitrogen compounds like HNO$_3$) are also photodissociated from solar UV radiation. The main effect of this sunlight dependence on ozone levels can be seen clearly in the results of~\citet{thomas2005} for the cases of GRB impacts that are centred over the Earth's northern or southern poles. Ozone depletion levels are consistently higher when the impact occurs at the beginning of or during polar nights (periods of extended darkness) compared to when the impact occurs during polar days (periods of extended sunlight). This effect is less prominent for ozone depletion around lower latitudes, since the seasonal variations in sunlight are more constant in these regions. The results of~\citet{thomas2005} also predict a small ($3 - 4$ per cent), short-term increase in the ozone depletion for GRB impacts that occur at midnight compared to at noon. This effect is relevant for GRBs with relatively short burst durations, of much less than one day. However, since the time-scale of energy deposition by the haloes considered in this work is significantly longer ($\sim 10$ days), we expect no differences between day/night time impacts.

Another important effect on total ozone abundance is atmospheric transport, which is believed to carry ozone and other stratospheric species from the tropics to the poles, likely due to air currents described by the Brewster-Dobson model~\citep{mohanakumar2008}. The transport of ozone and ozone-destroying compounds like NO$_x$ thus has an impact on ozone abundance primarily over different latitudes. This leads to ozone depletion that occurs in high latitudes or polar regions remaining local to that region, since there is minimal transport across the tropics. On the other hand, ozone depletion that occurs around the equator is `spread' polewards, which leads to depleted levels of ozone in other latitudes as well. The results of~\citet{thomas2005} confirm this effect, showing higher globally-averaged ozone depletion for GRB impacts centred on the equator compared to the poles, and depletion localised to each hemisphere for GRBs impacting at high latitudes. It seems that the isolation of ozone depletion to certain latitudes or hemispheres is more pronounced for phenomena like GRBs, which deposit energy into the atmosphere in relatively narrow beams. We expect the ozone depletion caused by nearby DM haloes to be more widespread over the globe, given the larger apparent area of energy deposition, similarly to the case of a nearby SN impact~\citep{thomas2018SN}.

The final noteworthy geographic effect is the change in ozone levels at varying altitudes. The destruction of ozone molecules in the upper layers of the stratosphere, due to an incident flux of gamma rays, allows solar UV radiation that would usually be absorbed to penetrate to the lower layers of the stratosphere ($\sim 15$ km). As solar UV is vital to the production of ozone, there is a slight increase in the abundance of ozone at these lower altitudes. This actually counters the overall depletion of ozone, and is accounted for in globally-averaged results from the GSFC model.

\subsection{Biological and environmental implications}\label{sec:biological_implications}

A common method of estimating the biological impact from an astrophysical ionizing event is with the use of Biological Weighting Functions (BWFs). These empirical functions quantify the amount of a particular type of biological damage from a specific type of radiation, often relative to some reference value. BWFs greatly simplify the calculation of intrinsically complex processes, such as DNA damage, and are useful tools for finding the biological damage from exposure to UV radiation. However, their effectiveness is dependent on the accurate determination of the irradiation of harmful UV reaching the surface. This has been calculated in relation to ozone depletion in various distinct ways in the past: by the use of a simplified Beer-Lambert-type expression~\citep{thomas2005}, by calculations that include both scattering and absorption effects~\citep{madronich1993}, and by the use of full radiative transfer models~\citep{thomas2015, madronich1998, mckenzie2007}. A detailed comparison between these methods was given in~\citet{thomas2015}, which found that using a Beer-Lambert-type calculation can overestimate the irradiance of UV-B on the surface by up to a factor of 2. When combined with a simple BWF~\citep{setlow1974} for UV-induced skin cancer in humans however, the simplified calculation yielded relative DNA damage factors of a similar magnitude to those found using the full radiative transfer model. This seems to indicate that simplified estimates of UV-B transmission through the atmosphere are valuable for gauging the overall plausibility of biological damage from ozone depletion, while not being as computationally expensive as more accurate methods. 

The holistic ecological impact from ozone depletion cannot be accurately determined with a single BWF; there are various distinct forms of biological damage, and various bands of solar radiation that are harmful to different organisms. While many previous studies focus on the increase of solar UV-B radiation (280 - 315 nm) due to ozone depletion, there is evidence that increased levels of other radiation bands, particularly UV-A (315 - 400 nm) and photosynthetically active radiation (PAR) (400 - 700 nm), can also have a significant effect on ecology~\citep{thomas2015}. Radiation exposure at these wavelengths can lead to a host of biological and ecological consequences, including erythema, cancer formation and inhibition of photosynthesis. These effects (and others) were modelled with separate BWFs in~\citet{thomas2015}, for the case of ozone depletion from a nearby GRB of 100 $\mathrm{kJ\,m^{-2}}$ fluence. While the results therein display large variability between each biological effect, they show significant damage to all levels of organism complexity. A particularly interesting result from this study is the effect on the UV Index (UVI), a numerical measure of the risk of skin damage in humans when exposed to UV radiation, adopted by the World Health Organisation in 1994~\citep{who1994}. The maximum value of the UVI after the GRB impact was as much as 29, with many regions having values greater than 11 -- the reference value for `extreme' risk in the UVI scale. These results provide strong evidence that ozone depletion from ionizing radiation would have a notable effect on the ecosystem, as well as affecting more than just primary producers. Since the atmospheric effects modelled here should correspond to those from ozone depletion caused by a GRB, we expect the gamma rays from DM haloes to produce similar ecological consequences to the ones presented in~\citet{thomas2015}. 

\section{Results and Discussion}\label{sec:results}

\subsection{Gamma-ray flux}\label{sec:results_fluxes}
\begin{figure*} 
\centering
\includegraphics[width=0.9\textwidth]{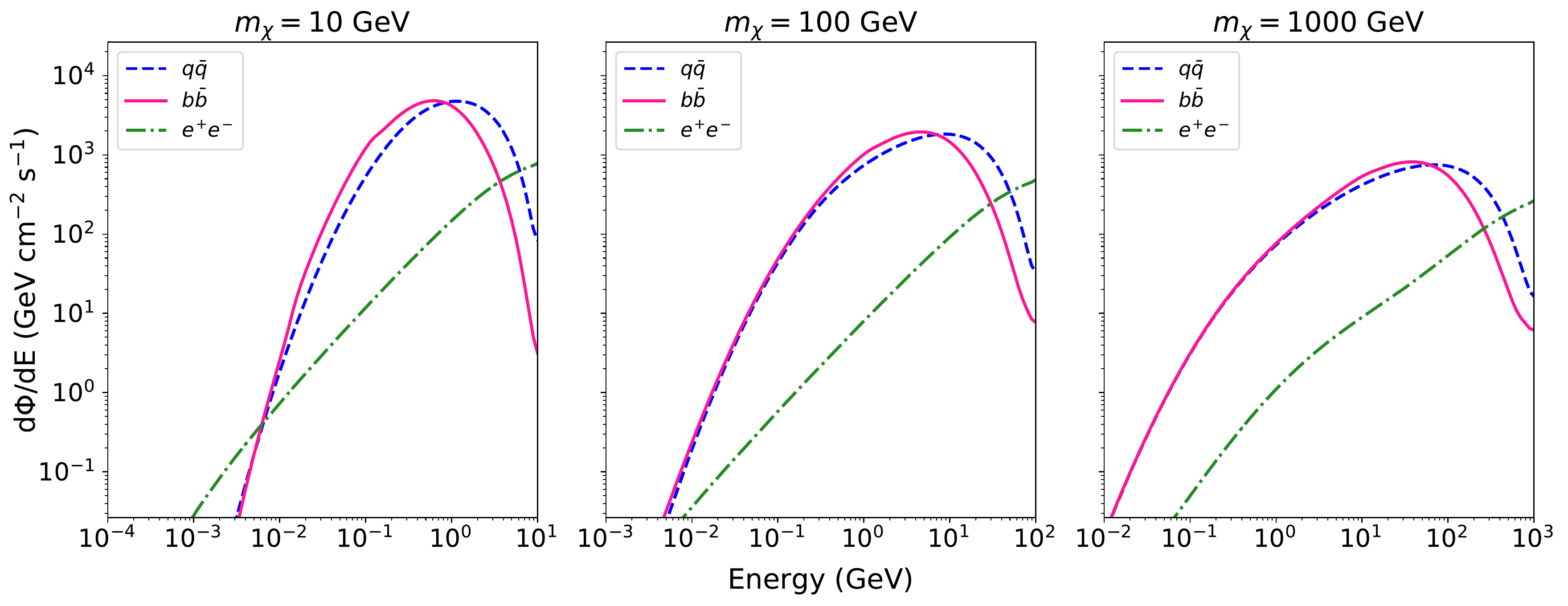}
\caption{Computed gamma-ray spectra of an UCMH containing WIMPs of mass 10, 100 and 1000 GeV annihilating into gamma-ray photons through the $q\bar{q}$, $b\bar{b}$ and $e^+e^-$ channels.}
\label{fig:ucmh_ann}
\end{figure*}

\begin{figure*} 
\centering
\includegraphics[width=0.9\textwidth]{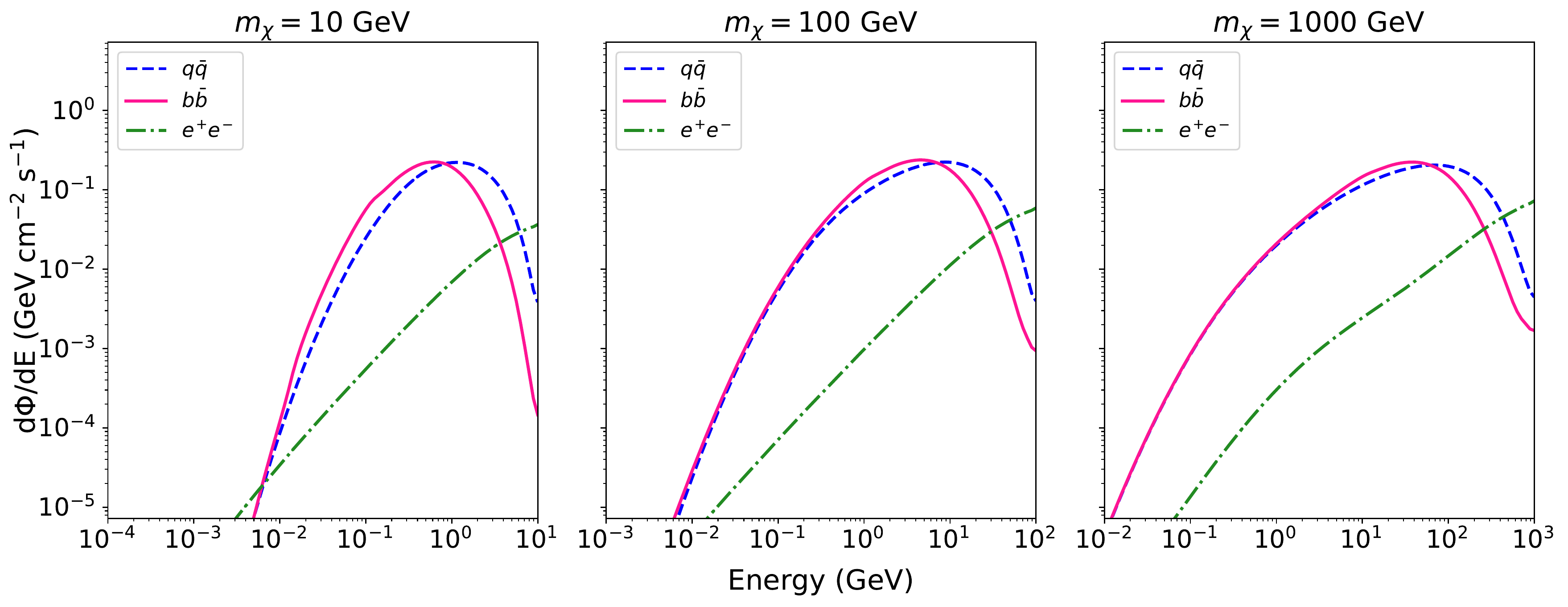}
\caption{Computed gamma-ray spectra of an UCMH containing WIMPs of mass 10, 100 and 1000 GeV decaying into gamma-ray photons through the $q\bar{q}$, $b\bar{b}$ and $e^+e^-$ channels.}
\label{fig:ucmh_dec}
\end{figure*}

\begin{figure*} 
\centering
\includegraphics[width=0.9\textwidth]{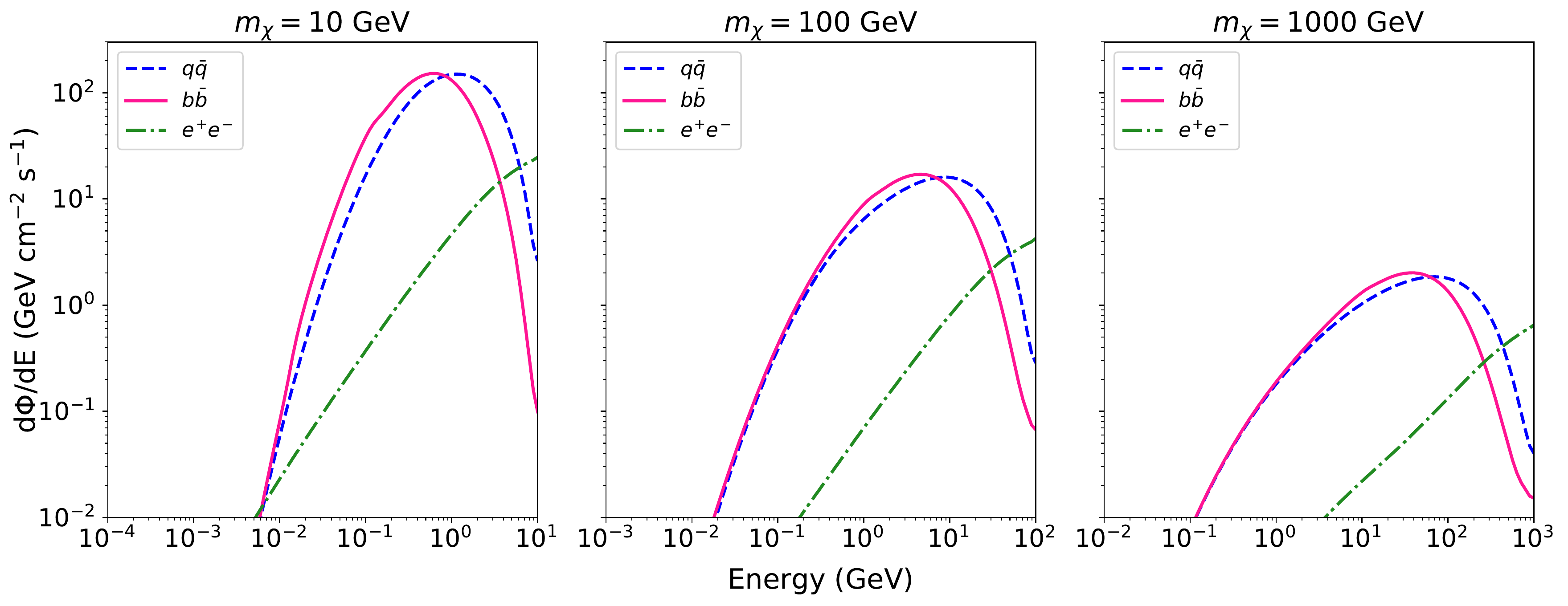}
\caption{Computed gamma-ray spectra of a Moore-like halo containing WIMPs of 10, 100 and 1000 GeV annihilating into gamma-ray photons through the $q\bar{q}$, $b\bar{b}$ and $e^+e^-$ channels.}
\label{fig:moore_ann}
\end{figure*}

The following plots show the produced instantaneous gamma-ray fluxes for haloes at a fixed distance of 1000 AU and a fixed total mass of 100 $\mathrm{M_{\odot}}$. We use the thermal relic WIMP annihilation cross-section of $\langle\sigma v\rangle = 3.0\times10^{-26} \, \mathrm{cm^3\,s^{-1}}$~\citep{cross-section} and decay rate of $\Gamma_{\mathrm{dec}} = 4.0\times10^{-26}\,\mathrm{s^{-1}}$~\citep{decay-rate} in all presented calculations. Fig.~\ref{fig:ucmh_ann} and Fig.~\ref{fig:ucmh_dec} show the differences in computed flux magnitudes between annihilation and decay channels (respectively) for an UCMH profile, whereas Fig.~\ref{fig:moore_ann} shows the flux magnitudes of annihilation channels for a Moore-like halo profile. These plots are shown to highlight the differences in the produced flux for the different halo profiles and annihilation/decay modes, which can easily be noted by the changes in the scale of the vertical axes. 

From these results, we note the following comparisons between the cases illustrated above. Firstly, the flux produced through the annihilation channels is several orders of magnitude larger than the flux produced through the corresponding decay channels. This is caused by the different dependence on WIMP density within the $J$ and $D$ factors used to calculate the flux, and is especially pronounced because of the high central densities of the haloes considered here. Secondly, we see that UCMHs produce significantly more flux than corresponding Moore-like haloes with the same mass and distance from the Earth. This is due primarily to the lower density of WIMPs within a Moore-like halo and the strong dependence on WIMP density in the flux calculation. There is also a contribution from the extension of the halo to this difference; the density profile of an UCMH has a steeper radial dependence than a Moore-like halo, so the radius within which 95 per cent  of the flux is produced is smaller for an UCMH than it is for a Moore-like halo. This leads to a reduction in the total flux received at the Earth by the larger halo when compared to the more compact UCMH. Finally, we note that the $e^+e^-$ intermediate channel produces less flux for most of the energy range considered when compared to the $q\bar{q}$ and $b\bar{b}$ channels. We have used these three intermediate channels to be representative of the larger set of possible channels. According to the results of~\citet{cirelli2011}, the $e^+e^-$ channel produces comparable fluxes to other leptonic channels, while the $q\bar{q}$, $b\bar{b}$ channels are comparable to other channels involving $W$ and $Z$ bosons and top quarks. The comparisons drawn above motivate our use of parameters in all of the remaining results. We have thus focused on UCMHs (with some comparisons to Moore-like haloes) and only consider annihilation of WIMPs through the $q\bar{q}$ channel. Another aspect of these results we would like to discuss is the dependence on the WIMP mass $m_{\chi}$ seen for each case. In general, the flux depends on the WIMP mass as $m_{\chi}^{-2}$ for annihilation and $m_{\chi}^{-1}$ for decay. In the central regions of the halo that reach the maximum density $\rho_{\mathrm{max}}$ however, the dependence on $m_{\chi}$ is cancelled out. Since UCMHs have a larger central region with $\rho = \rho_{\mathrm{max}}$ than Moore-like haloes because of their steep density profile, less of the halo is subject to the dependence on $m_{\chi}$. This explains why the fluxes produced from UCMHs show less variability with differing WIMP masses. 

\subsection{Encounter distances and rates}\label{sec:results_encounters}
\begin{figure*} 
\centering
\includegraphics[width=0.5\textwidth]{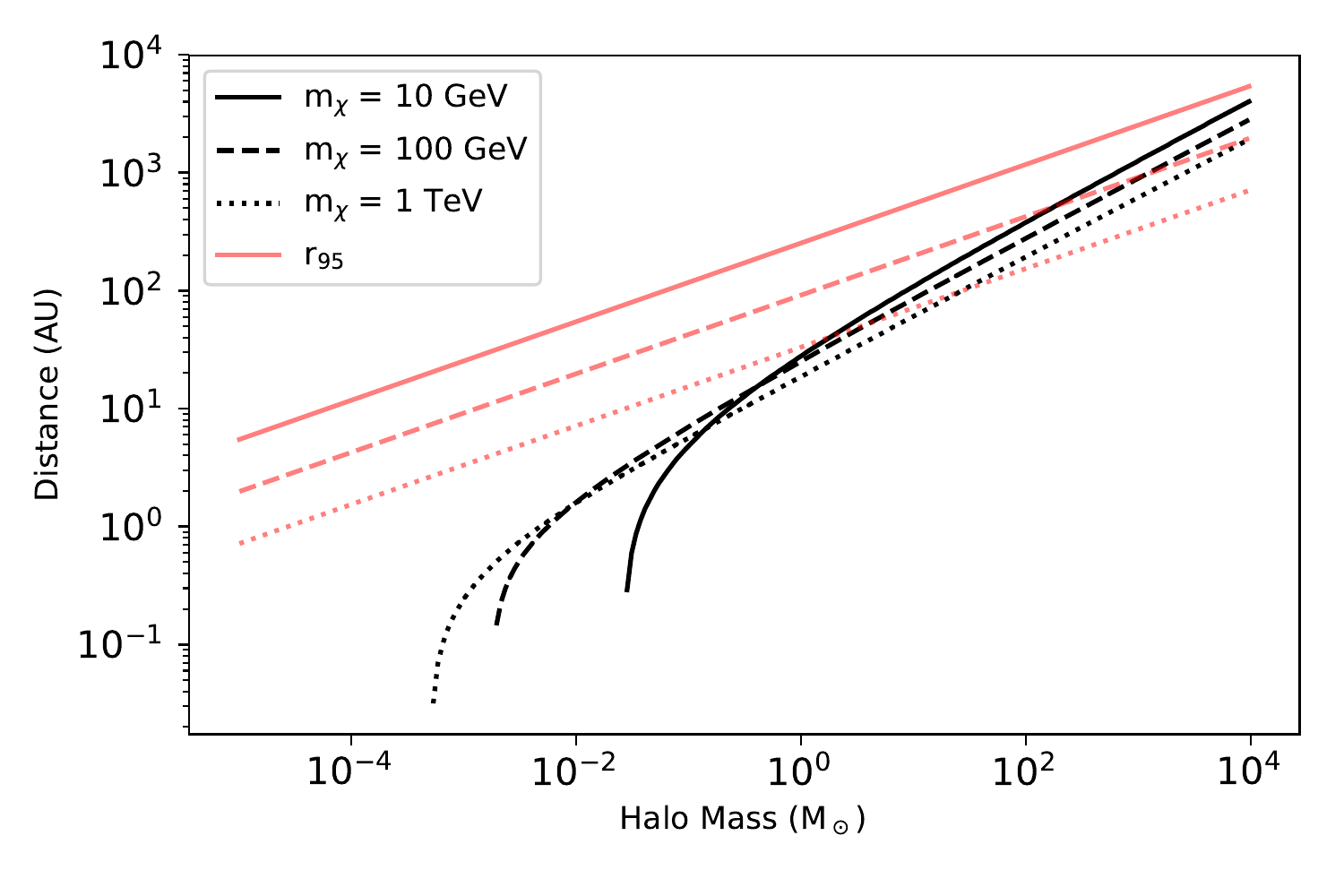}\includegraphics[width=0.5\textwidth]{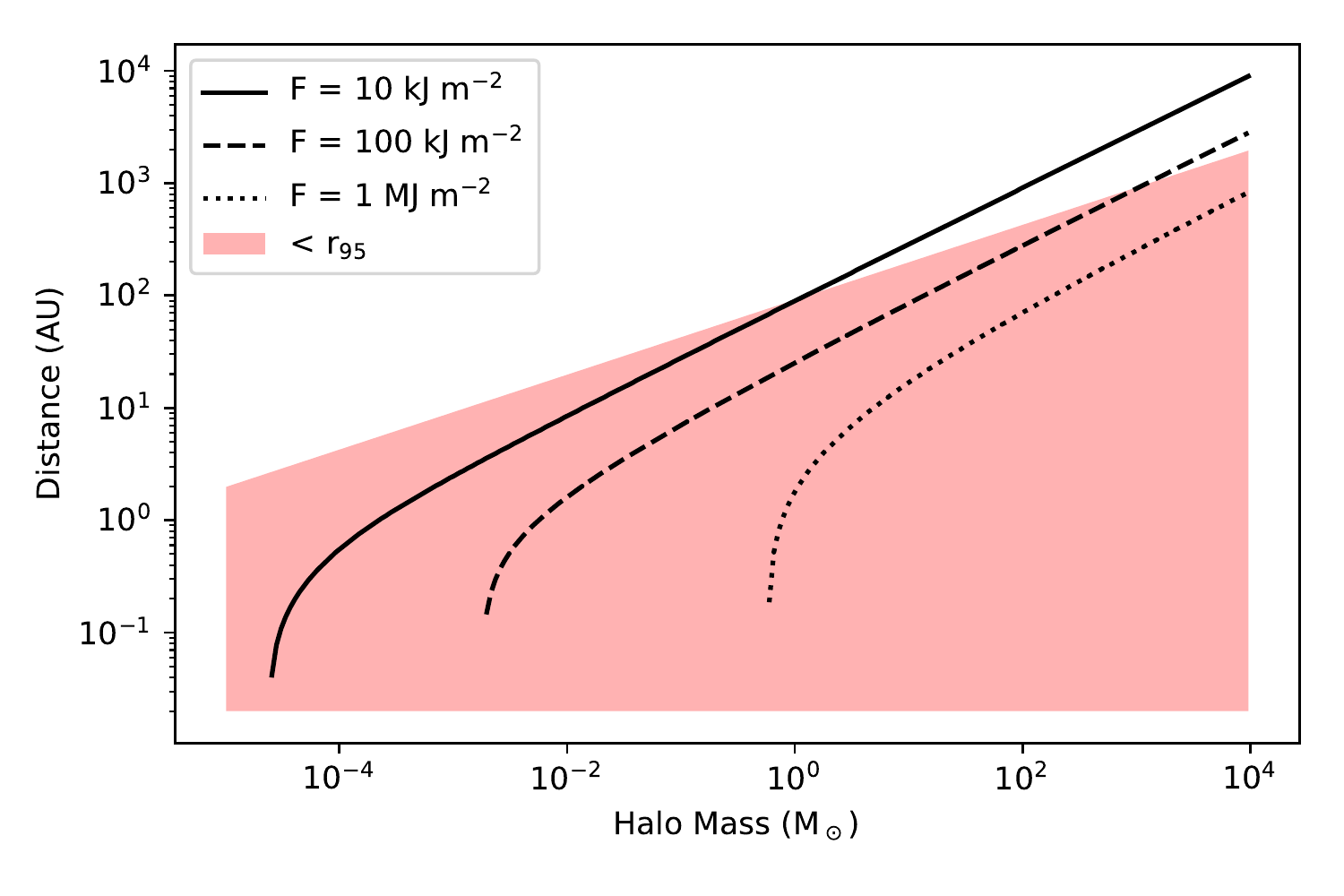}
\vspace{-0.2cm}
\caption{Encounter distances (in AU) for UCMHs containing WIMPs annihilating into gamma rays through the $q\bar{q}$ channel. Left: encounter distances calculated using a fixed fluence threshold of $F = 100 \, \mathrm{kJ\,m^{-2}}$ for a range of WIMP masses. Right: encounter distances calculated using a fixed WIMP mass of $m_{\chi} = 100$ GeV for a range of fluence thresholds.}
\label{fig:ucmhs100r95}
\end{figure*}

\begin{figure*} 
\centering
\includegraphics[width=0.5\textwidth]{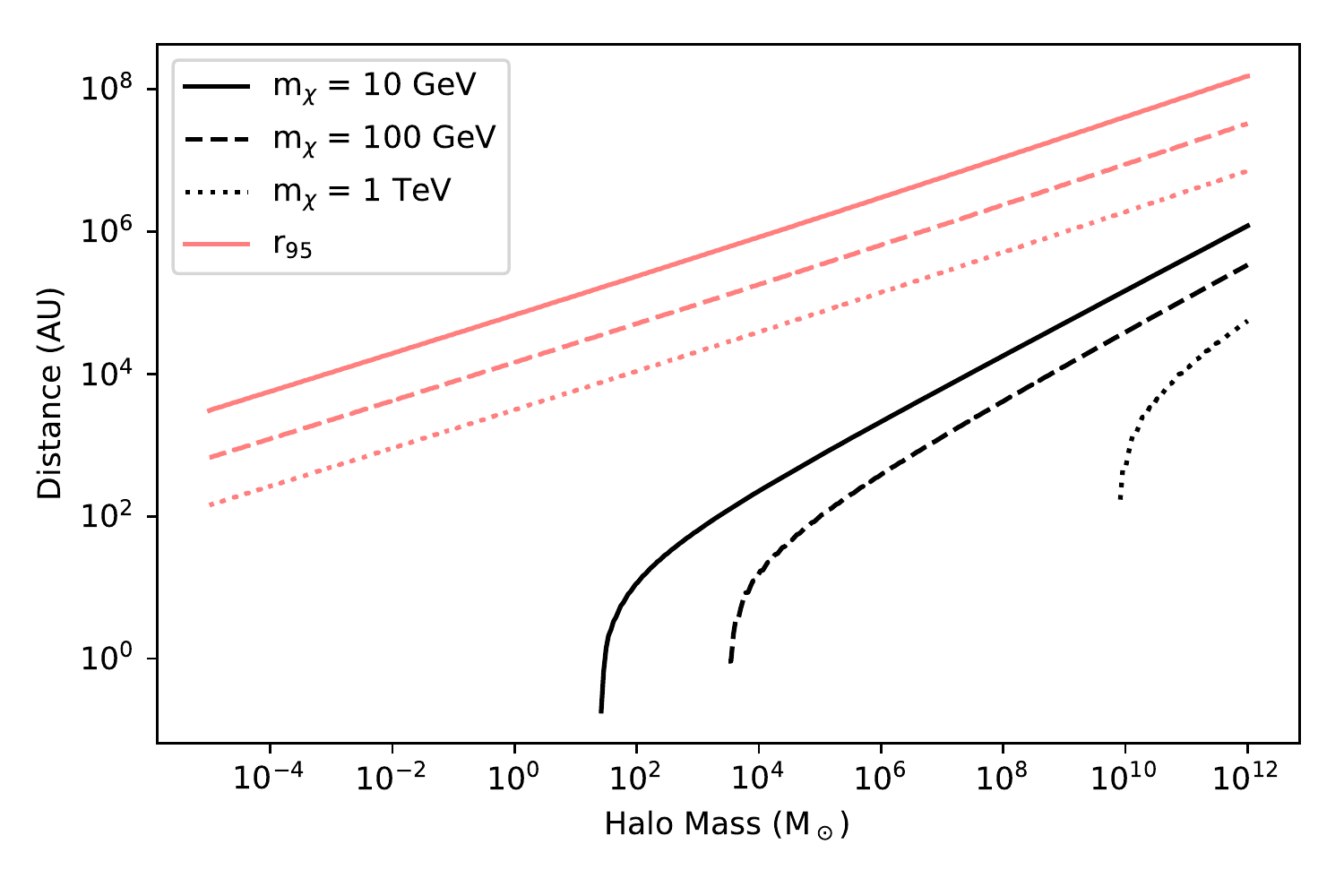}\includegraphics[width=0.5\textwidth]{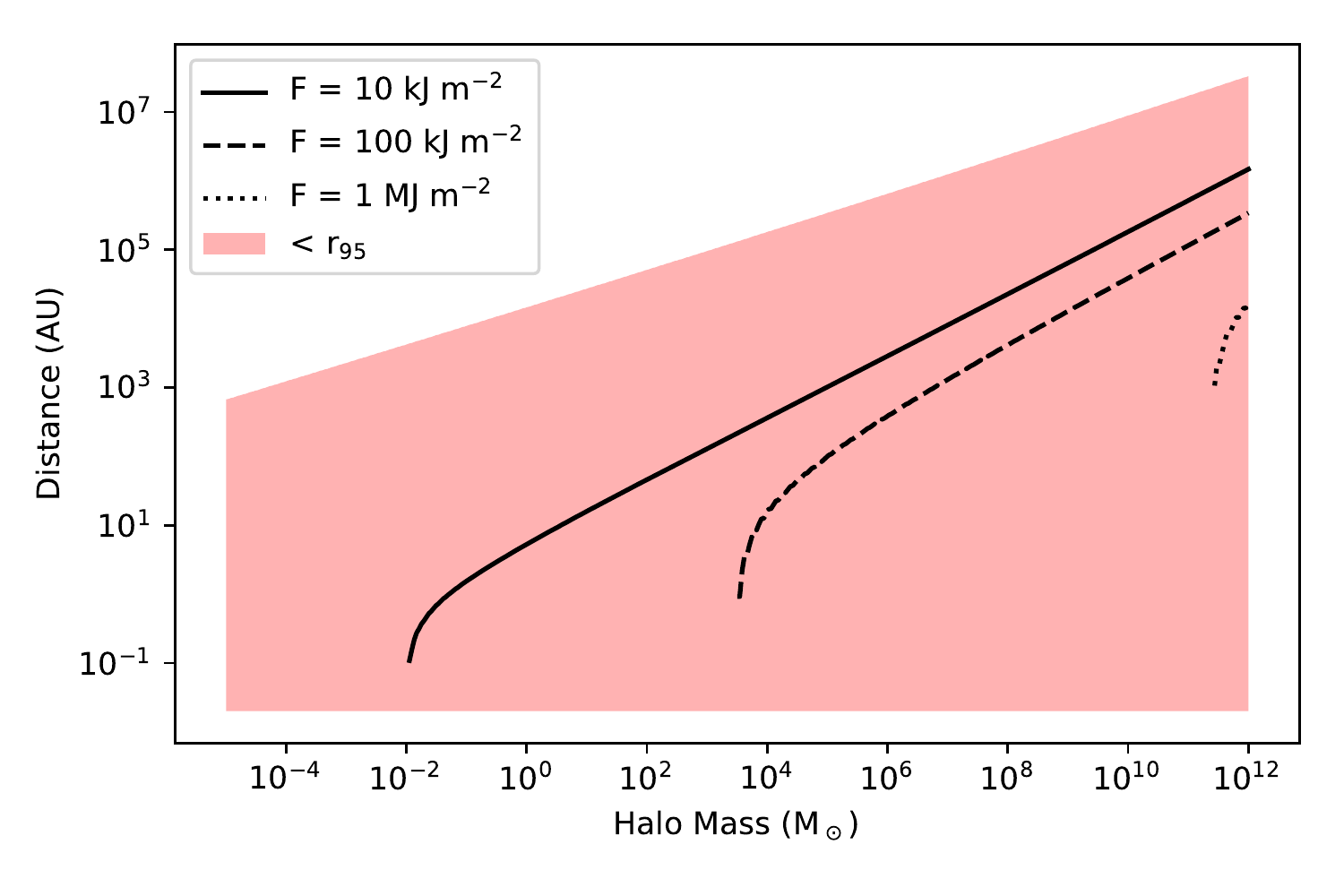}
\vspace{-0.2cm}
\caption{Encounter distances (in AU) for Moore-like haloes containing WIMPs annihilating into gamma rays through the $q\bar{q}$ channel. Left: encounter distances calculated using a fixed fluence threshold of $F = 100 \, \mathrm{kJ\,m^{-2}}$ for a range of WIMP masses. Right: encounter distances calculated using a fixed WIMP mass of $m_{\chi} = 100$ GeV for a range of fluence thresholds. The halo mass range displayed here has been extended compared to the one shown for UCMHs.}
\label{fig:moores100r95}
\end{figure*}

Fig.~\ref{fig:ucmhs100r95} and Fig.~\ref{fig:moores100r95} present the encounter distances ($d_{\mathrm{enc}}$) for each halo profile type, UCMH and Moore-like respectively, over a large range of possible halo masses. The encounter distances shown here are the maximum distances between each halo and the Earth within which a certain fluence threshold can be produced, and are found numerically from Equation~\ref{eqn:fluence}. The results are calculated using both a fixed fluence value for each WIMP mass and by having varying fluence thresholds for a fixed WIMP mass. Also shown in these figures are the values of $\mathrm{r_{95}}$, which represent the radii of each halo that contains $\sim 95$ per cent  of the total produced flux. In the cases where the distance between the halo and the Earth is less than the value of the radius r$_{95}$, which is seen with low-mass UCMHs and all Moore-like haloes, the Earth would have to reside within the halo to receive the required fluence threshold. 

These results show that the distances required to produce significant fluence thresholds are relatively low in all cases. Low-mass haloes especially would need to pass well within the solar system for this type of encounter, with some haloes not capable of producing the fluence threshold at all. We also note that Moore-like haloes have much lower encounter distances than UCMHs of the same mass, and the range of halo masses for Moore-like haloes shown in Fig.~\ref{fig:moores100r95} had to be extended to extremely large masses to fully compare the different halo profiles. 

An interesting feature of these results is that the Earth would have to reside within the halo whenever the value of $d_{\mathrm{enc}}$ is less than the halo's r$_{95}$ radius. This occurs for most low-mass UCMHs and for all Moore-like haloes regardless of WIMP mass or chosen fluence threshold. We expect this situation would cause other -- possibly catastrophic -- consequences for life on Earth, some of which have been described in Section~\ref{sec:intro}. One likely phenomenon in this scenario would be the gravitational disturbances of solar system bodies. Previous studies in this area~\citep{randall2014,kramer2016} have focused on the perturbations of comet orbits in the Oort cloud due to a dark disk of DM in the plane of the Milky Way. However, with the proximity and high density of the haloes considered in the encounters studied here, the gravitational disruptions are expected to extend to larger solar system bodies with possibly significant tidal effects. Other side effects of this type of encounter could be increased capture of WIMPs by the Earth's gravitational well, leading to heating of the core and an increase to flood-basalt volcanism~\citep{abbas1996,rampino2015}, and a higher level of carcinogenesis associated with WIMP-DNA collisions~\citep{collar1996,freese2012}. The high density of WIMPs within UCMHs/Moore-like haloes could accentuate these effects, leading to more severe threats to terrestrial biodiversity than the atmospheric changes considered here, however determining the full extent and likelihood of these effects is beyond the scope of this work. 

The total computed number of encounters experienced by the Earth ($\Gamma$) is shown in Table~\ref{tab:encounters}, for a range of fluence thresholds and mass distribution parameters $\sigma$ and $\mu$. Each subscript of $\Gamma$ represents the value of the fluence threshold used to determine the maximum encounter distance, i.e. 10, 100 or 1000 $\, \mathrm{kJ\,m^{-2}}$. The values in Table~\ref{tab:encounters} were calculated for the case of an UCMH density profile, and use a WIMP mass of 100 GeV and a time interval of $\tau = 4.5$ Gyr. These results also depend on the value of $f_{\text{H}}$, which is currently poorly constrained by observation. We have thus chosen to use the upper limit of the halo mass fraction, that all the galactic DM is contained within UCMHs, or $f_{\text{H}} = 1$. This limit is typically only allowed for distributions of very high or very low mass haloes, but as this parameter only enters into these calculations as a multiplicative factor the results can be scaled with more accurate constraints in the future. Similarly, we have chosen to neglect any relative motion between the Earth and UCMHs/Moore-like haloes in the value of $v_{\text{E}}$, which would in general depend on the orbital velocities of both objects around the galactic centre. There is strong evidence from numerical simulations that the stellar and galactic DM halo angular momentum profiles have no significant correlation~\citep{jiang2018,bullock2001spin}, which suggests that this relative motion is possible. Unfortunately, due to these null correlation results and the non-detection of halo substructure, there is a lack of precise measurements for halo properties like rotational velocity within the Milky Way halo. This forces us to neglect any relative motion and approximate $v_{\text{E}}$ as $\sim \,220 \,\mathrm{km \,s^{-1}}$, as in~\citet{green2016}. Like with the value of $f_{\text{H}}$ however, any future measurements of $v_{\text{E}}$ can be used to scale the results shown here.
 
\begin{table}
\caption{Number of encounters}
\label{tab:encounters}
\begin{tabular}{ c c c c c }
\hline
{$\sigma$} & {$\mu$ (M$\odot$)} & {$\Gamma_{10}$} & {$\Gamma_{100}$} & {$\Gamma_{1000}$}\\
\hline
\multirow{6}{*}{0.25} & $10^{-6}$ & $9.027\times10^{-43}$ & $ 0 $ & $ 0 $ \\
 & $10^{-4}$ & $5.590\times10^{-3}$ & $1.675\times 10^{-36} $ & $ 0 $ \\
 & $10^{-2}$ & $1.326\times10^{-2}$ & $4.760\times 10^{-4} $ & $2.697 \times10^{-67}$ \\
 & $1$  & $1.485\times10^{-2}$ & $1.173\times 10^{-3} $  & $6.313 \times10^{-6}$\\
 & $10^{2}$ & $1.532\times10^{-2}$ & $1.427\times 10^{-3} $ & $9.143 \times10^{-5}$\\
 & $10^{4}$ & $1.542\times10^{-2}$ & $1.507\times 10^{-3} $ & $1.330 \times10^{-4}$\\
\hline
\multirow{6}{*}{0.5} & $10^{-6}$ & $1.475\times10^{-14}$ & $3.311\times 10^{-56} $ & $ 0 $ \\
 & $10^{-4}$ & $5.499\times10^{-3}$ & $7.509\times 10^{-14} $ & $1.300\times 10^{-74}$ \\
 & $10^{-2}$ & $1.324\times10^{-2}$ & $4.709\times 10^{-4} $  & $7.489\times 10^{-23}$ \\
 & $1$ & $1.482\times10^{-2}$ & $1.171\times 10^{-3} $ & $7.538\times 10^{-6}$\\
 & $10^{2}$ & $1.530\times10^{-2}$  & $1.427\times 10^{-3} $ & $9.122\times 10^{-5}$\\
 & $10^{4}$ & $1.542\times10^{-2}$ & $1.508\times 10^{-3} $ & $1.329\times 10^{-4}$\\ 
\hline
\end{tabular}
\end{table}

The trend of these results can be understood as follows. For low values of $\mu$, which represents the mean halo mass in the distribution $\Psi(M)$, most haloes have insufficient mass to reach the necessary fluence threshold regardless of how close they pass to the Earth. This leads to the null or insignificant encounter rates seen in Table~\ref{tab:encounters}. For large values of $\mu$ there are fewer haloes present in the distribution, as the combined mass of all haloes is capped by the total mass of DM in the galaxy. We thus expect haloes with higher mass to be more sparsely spread throughout the galaxy, which leads to an upper limit on the number of encounters with these haloes as the value of $\mu$ increases. This is seen in Table~\ref{tab:encounters} wherein the value of $\Gamma$ plateaus as $\mu$ is increased to larger values. We also note that by increasing the value of $\sigma$, which defines the width of the mass distribution peak, the number of encounters increases. This effect is prominent only for distributions with low mean values, as a larger distribution peak allows more haloes to have sufficient mass to reach the required fluence threshold. 

The results shown here can be compared to similar calculations performed by~\citet{piran2014} to determine the rate of various types of GRBs. As this study also made use of the results found in~\citet{thomas2005}, in that it considered GRBs that reach fluence threshold of 10, 100 and 1000 $\mathrm{kJ\,m^{-2}}$, we can draw a rough comparison between the likelihoods of GRB impacts and halo encounters of equal fluence. The pertinent results from~\citet{piran2014} are those which show the probability of GRB impacts over the past 5 Gyr, which roughly corresponds to the time interval used in this work. Those results show a much larger probability of GRB impact, particularly for long GRBs (LGRBs) and short GRBs (sGRBs), than those calculated for an UCMH encounter of equivalent fluence and from this crude comparison it seems the rate of UCMH encounters most closely matches the rate of low-luminosity GRB (\textit{ll}GRB) impacts. These results also give some credence to the suggestion of a GRB being the cause of the Late-Ordovician mass extinction event investigated in~\citet{melott2009}, with a 50 per cent  probability that a LGRB impact of 100 $\mathrm{kJ\,m^{-2}}$ has  occurred in the last 0.5 Gyr. LGRBs represent the overall most likely phenomenon for these impacts, with sGRBs, \textit{ll}GRBs and encounters with UCMHs all comparatively less likely. Finally, we also note that in general Moore-like haloes produce less fluence than UCMHs with equal mass and would need to pass closer to the Earth to reach these fluence thresholds. We thus expect Moore-like halo profiles to have an even lower rate of encounters than those shown here for UCMH profiles.

\subsection{Ozone depletion}\label{sec:results_ozone}

The primary aim of using the GSFC atmospheric model in this work was to determine the extent of ozone depletion in the atmosphere due to a incident flux of gamma rays that originate from WIMP annihilations. We have simulated six cases of DM haloes with various WIMP and halo parameters to be used as input for the GSFC model, each of which are summarised in Table~\ref{tab:ozonetab}. The gamma-ray fluence resulting from the UCMH cases have been normalised to approximately $100 \,\mathrm{kJ\,m^{-2}}$ by numerically adjusting the distance to each halo, with this value chosen to allow for easy comparison with other studies that consider it a reference fluence value. We have also included cases for Moore-like halo profiles, however the distances to these haloes are not normalised using a certain fluence threshold, and are instead taken to be equal to the distance of each UCMH of equivalent WIMP mass. Having the distances to UCMHs and Moore-like haloes fixed for each corresponding case allows us to compare the total ozone depletion caused by the two profiles as directly as possible. We also keep the total mass of each halo at a fixed value of 100 $\mathrm{M_{\odot}}$, with a total duration of energy deposition of 10 days. The choice of time interval was motivated by two factors. Firstly, the instantaneous flux from UCMH haloes drops rapidly as the halo moves through its orbital trajectory, such that time intervals higher than $\sim$ 10 days lead to a negligible increase in flux output. This is due to the inverse square factor of distance in the $J$ and $D$ factors and the output flux which is strongly peaked around the center of the halo. Secondly, the results of~\citet{ejzak2007} indicate that there is approximately no change in ozone depletion levels for any variation in energy deposition time, in the range of $0.1-10^8$ s, given a constant fluence. Since we are considering a fixed fluence value of 100 $\mathrm{kJ\,m^{-2}}$, the value used here of 10 days provides a reasonable estimate of the time-scale of this effect while keeping the ozone depletion levels independent of any time interval variations. 

\begin{figure*}\label{fig:ozone}
\centering
\includegraphics[width=\textwidth]{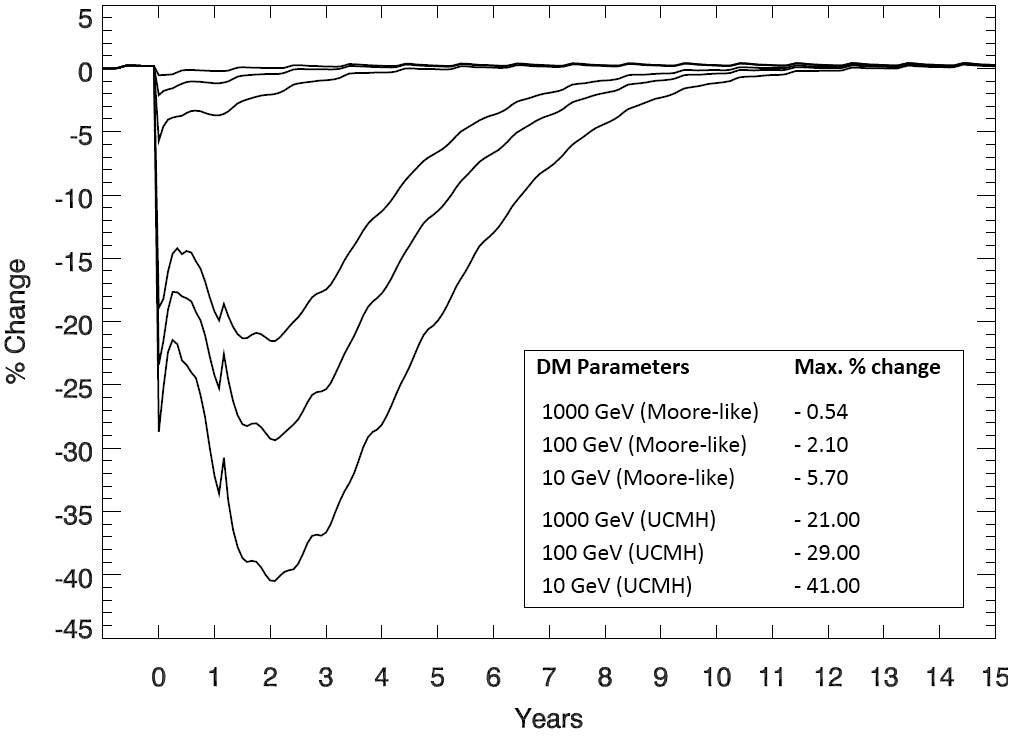}
\caption{Relative changes in the globally averaged ozone column density over time. The injection of ionizing gamma rays into the atmosphere begins at time 0, and lasts for 10 days. Separate curves represent results for individual DM haloes with different halo and WIMP properties, which are summarised in Table~\ref{tab:ozonetab}. In each case the ionization is centred on the Earth's equator, and occurs during the March equinox.}
\end{figure*}

The gamma-ray spectra resulting from the UCMH or Moore-like halo in each case discussed above was used to determine the ionization profiles, which were then used as input for the atmospheric model. In all cases shown here we set the encounters to occur during late March (roughly coinciding with the March equinox) and over the Earth's equator. The results of these simulations are shown in Fig.~\ref{fig:ozone}, which displays the percentage change in the globally averaged column density of ozone in the atmosphere. 

\begin{table}
\caption{Halo and WIMP parameter selection}
\label{tab:ozonetab}
\begin{tabular}{ c c c c c }
\hline
{Halo profile} & {m$_{\chi}$ (GeV)} & {$\delta$ ($^{\circ}$)} & {$F$ ($\mathrm{kJ\,m^{-2}}$)} & {Max. \% change}\\
\hline
\multirow{3}{*}{UCMH} & 10 & 104 & 100.25 & -41.00\\
 & 100 & 68 & 100.76 & -29.00\\
 & 1000 & 38 & 100.75 & -21.00\\
\hline
\multirow{3}{*}{Moore-like} & 10 & 179 & 0.64 & -5.70\\
 & 100 & 177 & 0.16 & -2.10\\
 & 1000 & 174 & 0.041 & -0.54\\
\hline
\end{tabular}
\end{table}

We make note of several general trends in the results shown in Fig.~\ref{fig:ozone}. As expected, the ozone depletion from Moore-like haloes is significantly lower than from UCMHs with equivalent mass WIMPs due to their lower flux output, which is a direct consequence of their shallower density profiles. There is also a dependence on the WIMP mass for each halo type, with lower WIMP masses producing larger levels of ozone depletion. This relationship results primarily from two effects, involving the central density of the halo and the ionization profile. Firstly, the area of energy deposition into the atmosphere ($\delta$) is dependent on the radius ($r_{95}$) of the halo, with larger radii haloes thus causing more globally-averaged ozone depletion. Since the value of $r_{95}$ is determined by the extent of the dense of central region of each halo, which in turn is dependent on the WIMP mass, haloes with larger radii are those which contain lighter WIMPs. Secondly, the ionization profile -- the input for the GSFC model -- favours lower energies in the range that we have considered. As the peak of the output flux spectra is shifted to lower energies for lower WIMP masses, these lower mass WIMPs produce a larger amount of ionization at these energies. The ionization profile directly changes the number of produced NO and OH molecules in the atmosphere and thus affects the total level of ozone depletion. The combination of these two effects form the basis of the relationship observed between ozone depletion and WIMP mass. 

The depletion of ozone from UCMHs with a fluence of 100 $\mathrm{kJ\,m^{-2}}$ varies between -21 and -41 per cent, which is comparable to the depletion of ozone from GRBs with the same fluence. For instance, in~\citet{thomas2005}, a GRB of 100 $\mathrm{kJ\,m^{-2}}$ fluence occurring during the March equinox led to -36 per cent ozone depletion, which is within the range of values found here. The ozone recovery time of $\sim$ 10 years seen here is also similar to the value found for GRBs, and these results suggest that the spectrum of gamma rays resulting from WIMP annihilations in an UCMH would lead to similar atmospheric effects as those from a GRB. We expect the biological effects from this event to mimic those discussed for example in~\citet{melott2009}, which are also shown to be compatible with the fossil record from the Late-Ordovician mass extinction event. Any variations to the geographic and seasonal input parameters of the model should lead to results consistent with the considerations outlined in Section~\ref{sec:atmospheric_modelling}.

\section{Conclusion}\label{sec:conclusion}

We have calculated the predicted gamma-ray annihilation fluxes from a set of compact DM haloes that pass nearby the solar system, and simulated the resulting changes to ozone abundance in the Earth's atmosphere. We find large variation in the magnitude of ozone depletion, due to the range of the allowed parameter spaces of WIMPs and DM haloes, with our most extreme case reaching globally-averaged depletion levels of -41 per cent. As there is evidence that the atmospheric and ecological effects of ionizing radiation in the atmosphere are applicable to different astrophysical sources, this work can be compared to previous results in the literature regarding terrestrial effects of GRBs and SN. High levels of ozone depletion, and the resulting increase in solar UV radiation penetrating the atmosphere, have been shown to produce a significant amount of biological damage in a wide range of living organisms. These biological effects have been linked to major changes in the biodiversity of life on Earth in the past, including the Late-Ordovician mass extinction event. These results suggest that compact DM haloes composed of WIMPs could act as another astrophysical source of ozone depletion, possibly affecting the history of biodiversity on Earth. 

We note that several halo and WIMP configurations lead to larger gamma-ray fluxes, which increases the extent and likelihood of ozone depletion in the atmosphere. UCMHs, due to their steep density profiles, produce a higher flux than Moore-like haloes of corresponding mass and distance. WIMP annihilation, rather than decay, is also the dominant mechanism of gamma-ray production in these haloes, with flux magnitudes several orders of magnitude larger. There is also a strong dependence on the WIMP mass in the results, with larger masses resulting in lower fluxes and less ozone depletion. This was attributed to the presence of the WIMP mass in the calculation of the core radius of each halo and in the ionization profile used as input to the GSFC model. 

Finally, we note that the estimated rate of encounters of this effect is relatively low, for all considered cases. This is due primarily to the proximity of the haloes, as they need to be extremely close to or even within the solar system to generate a sufficient gamma-ray flux. The highest probability of an encounter occurs when the peak of the halo mass distribution is shifted to very large masses, and becomes negligible when the mean halo mass is relatively small. Some parameters in this calculation, like the halo mass fraction and relative orbital velocity, are currently poorly constrained and their estimation leads to uncertainty in the final number of encounters. However, the results presented here can be easily scaled with any future limits on these values. 

While we have shown that UCMHs composed of WIMPs are capable of depleting a significant amount of atmospheric ozone, their estimated distribution in the galaxy makes the likelihood of such an event relatively low. In the cases of haloes with Moore-like density profiles, WIMP decay or leptonic annihilation channels, this likelihood is further decreased. The close proximity of these haloes also implies that other life-threatening effects would be prominent in such an encounter, such as orbital disruptions, an increased level of comet impacts, flood-basalt volcanism and widespread carcinogenesis from direct collisions with WIMPs. We conclude that compact DM haloes can act as a source astrophysical ionizing radiation capable of affecting the atmosphere and biodiversity on Earth, but phenomena such as GRBs and SN seem a more likely source of these effects.

\section*{Acknowledgements}

This work is based on the research that was supported by the National Research Foundation of South Africa (Bursary No 112332). G.B acknowledges support from a National Research Foundation of South Africa Thuthuka grant no. 117969.

\section*{Data Availability}

The data underlying this article are available in the article and in its online supplementary material.


\bibliographystyle{mnras}
\bibliography{references} 

\bsp	
\label{lastpage}
\end{document}